\documentclass[aps,onecolumn, superscriptaddress]{revtex4}
\input epsf
\usepackage[dvips]{graphicx}
\usepackage{amsmath,amssymb,latexsym}
\newcommand{\nablada}{ \bar{\nabla}_{a} }
\newcommand{\nablaua}{ \bar{\nabla}^{a} }
\newcommand{\nabladb}{ \bar{\nabla}_{b} }
\newcommand{\nablaub}{ \bar{\nabla}^{b} }
\newcommand{\nabladc}{ \bar{\nabla}_{c} }
\newcommand{\nablauc}{ \bar{\nabla}^{c} }

\newcommand{\nabladm}{ \bar{\nabla}_{m} }
\newcommand{\nablaum}{ \bar{\nabla}^{m} }
\newcommand{\nabladn}{ \bar{\nabla}_{n} }

\newcommand{\nabladell}{ \bar{\nabla}_{\ell} }

\newcommand{\nabladi}{ \bar{\nabla}_{i} }

\newcommand{\deltapiuab}{ \delta \pi^{ab} }
\newcommand{\deltapidab}{ \delta \pi_{ab} }
\newcommand{\deltapitr}{ \delta \pi }

\newcommand{\gb}{\bar{g}}

\newcommand{\sqrtg}{ \sqrt{|g|} }
\newcommand{\sqrtgb}{\sqrt{| \bar{g}| } }

\newcommand{ \Deltabar }{ \bar{\Delta} }

\newcommand{ \phibdot }{ \dot{ \bar{\phi} } }

\newcommand{\bea}{\begin{eqnarray}}
\newcommand{\eea}{\end{eqnarray}}

\begin{document}

\title{ On the initial value problem for second order scalar fluctuations in Einstein static  }
\author{ B.~Losic }
\affiliation{ Department of Physics \& Astronomy, 
University of British Columbia
6224 Agricultural Road
Vancouver, B.C. V6T 1Z1
Canada }
\author{ W.G.~Unruh }
\affiliation{ Canadian Institute for Advanced Research, Cosmology and Gravitation Program  }
\email{ blosic@physics.ubc.ca   ;  unruh@physics.ubc.ca}

\date{April 5, 2004 }

\begin{abstract}
We consider fluctuations in a perfect irrotational fluid coupled to gravity in an Einstein static universe background. 
We show that the homogeneous linear perturbations of the scalar and metric fluctuations in the Einstein static universe must be
present if the second order constraint equations are to be integrable. I.e., the 'linearization stability' constraint forces the 
presence of these homogeneous modes. Since these linear homogeneous scalar modes are well known to be exponentially unstable, the 
tactic of neglecting these modes to create a long-lived, almost Einstein universe does not work, even if all higher order (L $>$ 1) 
modes are dynamically stable. 

\end{abstract}

\maketitle
\section{Introduction}

Recently Barrow et al \cite{Barrow:2003ni} re-examined the stability of the Einstein static spacetime against arbitrary linear fluctuations
in the metric and matter and found, surprisingly, that some modes are stable. More precisely, they found that given a sufficiently large 
speed of sound in the background, all non-gauge inhomogeneous scalar modes were neutrally stable (i.e. the fluctuations are not damped), and furthermore
vector and tensor modes were neutrally stable on all scales irrespective of the equation of state in the background.
The essence of this stability arises from the spatial compactness of the Einstein static spacetime, i.e. there exists a maximum physical wavelength in this 
closed space and furthermore the Jeans length is a significant fraction of this maximum scale. It turns out 
that for specific equations of state in the background matter all physical wavelengths fall below the Jeans wavelength and the modes are 
thus stable for the usual reasons. 

This surprising non-Newtonian stability for a large class of fluctuations, which was pointed out earlier 
in different and more restricted contexts in \cite{Harrison:fb} and \cite{Gibbons}, is one of the key elements of support for the 
Emergent Universe models proposed recently by Ellis et al in \cite{Ellis:ba}. These models explicitly construct spatially closed, positively
curved, cosmologies which do not bounce and in which inflation (triggered by precisely the famous homogeneous instability 
of Einstein static) is not preceded by an era of deceleration, by contrast to deSitter and most other models of closed inflation. Such 
constructions have no initial singularity, undergo the usual inflationary period ending in the usual reheating era, and immediately solve 
the horizon problem owing to the staticity of the initial state. However another key feature of these models is that a finite number of 
e-foldings of inflation occurs over an {\it infinite} time in the past. It is critical for this scenario that the homogeneous mode, which is
always exponentially unstable in linear order, be excluded with high accuracy.

In the present paper we investigate the initial value problem for second order inhomogeneous fluctuations about the Einstein static solution, 
as made precise below. For some time now it has been known  (see, e.g., \cite{Brill:1973}) that the initial value equations of gravity have peculiar 
properties in a closed universe with symmetries. The crucial idea is that there are certain flux integrals, associated with those symmetries, which are
constrained to be zero. Thus not all solutions to the linear perturbations equations are actually perturbations of some exact solution to the Einstein equations. 
Technically, the tangent space defined by the linearized initial value constraints is actually larger than the manifold of solutions. Only those solutions 
to the linearized equations which obey an additional second order constraint can be the linearization of an actual family of solutions to the full Einstein
equations. This feature of the perturbations of the Einstein equations is  called ``linearization stabilty". It does not refer to the dynamic stability 
of pertubative solutions, but to the above ``stability" of the linearized solutions as genuine approximations to exact solutions of the
Einstein equations.


The nature of these higher order restrictions on the linear fluctuations about a certain background space, which are necessary conditions to ensure the 
so-called `linearization stability` of that background space ( see \cite{Moncrief:un}, \cite{Moncrief:cz} ), is such that for every 
Killing vector in the background there is an associated higher order constraint. 
In this paper we focus on the higher order constraint associated with the timelike Killing field of the Einstein static initial state. 


We find, in the case of a general irrotational perfect fluid with cosmological constant, that there are no nontrivial solutions to both 
the linearized constraints and this global nonlinear constraint when we exclude the linear homogeneous scalar metric and matter fluctuations. In other words, 
the leading order linear metric and matter fluctuations must be trivial if their linear seeds do not include `zero mode` homogeneous fluctuations. It is well 
known that these homogeneous modes are dynamically unstable for any (causal) equation state of the matter, just as all perturbations of an Einstein static 
dust model are unstable. This would seem to suggest that if the universe is in a neighbourhood of the Einstein static solution then it does {\it not} stay 
there, even if perturbed with the neutrally stable modes found in reference \cite{Barrow:2003ni}. This is one 
of the first instances where the linearization stability issues make a physical difference in the analysis of metric and matter stability 
(see also Brill, \cite{Brill:1973}).

The paper is organized as follows. In section II we briefly outline the details of the Einstein static background model. In section III we 
define the constraint equations in the standard ADM decomposition of the Einstein equations and also define an orthogonal decomposition 
(following  \cite{Brill:1973}) of perturbations into transverse and longitudinal parts. In Section IV we formulate and compute the 
nonlinear constraints, leaving conclusions for Section V. The entire analysis is quite similar to that in Brill and Deser's original paper 
\cite{Brill:1973}.

\section{ Einstein static spacetime }

Consider a FRW universe in comoving coordinates $(t,\vec{x})$ with scale factor $a(t)$, with signature (-1,1,1,1), and with a perfect fluid with energy density 
$\rho$ and pressure $p$. The equations of motion for the scale factor $a(t)$ are, according to the Einstein equations,
\begin{eqnarray}
\frac{ \ddot{a} }{a } &=& -\frac{ \kappa }{3} \left[ \rho (1 + 3 w) - \Lambda \right], \\
H^2 &=& \frac{  \kappa }{3} ( \rho + \Lambda)  - \frac{K}{a^2}, 
\end{eqnarray}
where $K \equiv \pm 1, 0$ is the constant curvature of the $t = const$ slices, $H \equiv \partial_{t} ln( a )$ is the Hubble parameter, $\Lambda$ is a cosmological constant,
$w \equiv \frac{p}{\rho}$, and $\kappa \equiv 8 \pi G$ in units where $c = 1$. One can combine these equations to form $\dot{H} = -\frac{\kappa}{2} (\rho(1+w)) + \frac{K}{a^2}$.

As Einstein found some time ago, demanding that the universe be static ($\dot{a} = \ddot{a} = 0$ ) obviously sets $K$ to be positive (take it to be 1) 
and leads to constraints relating the initial energy density and pressure of the fluid to $\Lambda$. The equilibrium radius of such a static universe  
is set by these constraints to be  
\begin{eqnarray}
a_{0}^2 = \frac{1+3w}{\Lambda (1+w)} ,
\end{eqnarray}
where $(1 + w)\rho > 0$. In the case of dust it is seen that $a_{0}^2 = 1 /  \Lambda$.

Consider the general perfect fluid equations. The equations of motion of the perfect fluid are given by the conservation equations of
$T^{\mu\nu}=(\rho+p)u^{\mu}u^{\nu} +pg^{\mu\nu}$, or
\bea
(\rho+p)_{;\mu}u^\mu u_\nu +(\rho+p)u^\mu_{;\mu}u_{\nu}+(\rho+p)u^\mu
u_{\nu;\mu}+ p_{;\nu}=0
\eea
Dotting this with $u^{\nu}$ we get the primary conservation law
\bea
\rho_{;\mu}u^\mu +(\rho+p)u^\mu_{;\mu}=0
\eea
Multiplying this by $u_\nu$ and subtracting  from the above equation we get
\bea
\nonumber
0&=&(\rho+p) u^\mu u_{\nu;\mu} + p_{;\nu}+p_{, \mu}u^\mu u_\nu \\
&=& ( \frac{ \rho + p}{\sqrt{N}} )( \sqrt{N} u_{ [\nu } )_{;\mu]} u^{\mu},
\eea
where
\bea
N(\rho)= exp\left(2\int {dp\over \rho+p}\right)
\eea
is a function of $\rho$ where we assume that there exists some constitutive equation $p=p(\rho)$. These equations are  trivially satisfied if we assume that 
the term $ N^{1\over 2}(\rho)u_\mu$ satisfies
\bea
N^{1\over 2}(\rho)u_\mu=\phi_{;\mu},
\eea
where $\phi$ is a scalar potential for the flow. This is an appropriate generalisation of the irrotational conditions of non-relativistic flow. Finally, 
multiplying this equation by itself we have that
\bea
N(\rho)={-\phi_{;\mu}\phi^{;\mu}}
\eea
which is the equivalent of Bernouli's equation for non-relativistic
irrotational flow. We will concern ourselves with constitutive equations where $p$ is linearly proportional to $\rho$.

We wish to treat the case of an arbitrary {\it irrotational} perfect fluid in terms of  
this scalar field $\phi$ that acts as the velocity potential of the fluid (as in \cite{Garriga:1999vw} ).  We choose the scalar field to have the action 
\begin{eqnarray}
S &=& \frac{1}{2\alpha} \int ( -\phi^{,\lambda} \phi_{, \lambda} )^{\alpha} \sqrt{-|g|} d^{4}x, \ \ \alpha \in \Re 
\end{eqnarray}
Comparing the stress energy that results from varying the above action 
with respect to the metric with that of a perfect fluid, we can easily identify the corresponding energy density and pressure in terms of $\phi$:
\begin{eqnarray}
\rho &=& ( \alpha - \frac{1}{2} ) N^{\alpha}, \\
p &=& \frac{ N^{\alpha} }{2}.  
\end{eqnarray}
For $\alpha = 1$ we obtain the stiff ($w = 1$)  perfect fluid, i.e. a minimally coupled scalar field. However, note 
that the speed of sound is given by $c_{s}^2 = p / \rho = ( 2 \alpha -1 )^{-1}$ and thus causality restricts $\alpha \ge 1$. It turns out that stable (inhomogeneous) scalar 
fluctuation modes only exist when the background speed of sound satisfies the bound $v_{s}^2 \ge \frac{1}{5}$ (\cite{Barrow:2003ni}, \cite{Gibbons}) for no Jeans 
instability, which translates into $\alpha \le 3$. I.e. there is no Jeans instability for any physical inhomogeneous modes given these conditions. Thus we take 
$1 \le \alpha \le 3$ in what follows. We also note in passing that in order for the four velocity to 
be timelike the gradients of the velocity potential $\phi$ are restricted to ones such that their temporal gradient dominates their spatial counterpart. In this 
model we take the background $\phi = ({\rho_0\over \alpha-1/2})^{1\over\alpha} t$ to satisfy equation (7) in the background, so the timelike condition will always be 
satisfied for the perturbations. 

In any case the generalized Einstein static initial conditions (in terms of $\phi$) become
\begin{eqnarray}
N^{\alpha} \frac{ \alpha + 1}{2} &=& \Lambda = \frac{ \alpha + 1}{\alpha \kappa a_{0}^2}, 
\end{eqnarray}
i.e. $a_{0}^2 = (\alpha + 1)/ (\alpha \Lambda)$ and $N = ( 2 \Lambda / (\kappa (\alpha + 1)) )^{\frac{1}{\alpha}}$.

\section{ Initial-value constraint equations  } 

The gravitational field may be characterized in terms of the initial three-geometry $g_{ij}$ and extrinsic curvature $K^{ij}$ of some spacelike surface. 
These twelve quantities may be rewritten into a more convenient hamiltonian form 
by defining the momentum density $\pi^{ij} \equiv \sqrt{ |g| } (K^{ij} - K^{\ell}_{\ \ell} g^{ij} )$, which is conjugate to $g_{ij}$. The 
phase space for Einstein's equations is then some suitable function space of pairs ($g_{ab}, \pi^{ab}$) over a three-dimensional manifold
$M$ (which we take as compact and without boundary), and the constraint subset of phase space is defined by the four 
initial-value constraints. The constraints are nonlinear and we expand them order by order in the strength of the metric and matter 
perturbations. Quantities with overbars will indicate zeroth order, background, quantities and a bar indicates covariant differentiation 
with respect to the background metric (used intechangably with the standard $\nabladi$ notation). 

The constraints with matter sources $(\rho, J^{a})$ are 
\begin{eqnarray}
- \frac{1}{ \sqrtg } \left( \pi^{ab} \pi_{ab} - \frac{1}{2} \pi^{2} \right) + \sqrtg {}^{(3)}R &=& 2 \kappa \sqrtg \rho, \\
( \pi^{ab} )_{\ |b} &=& \kappa J^{a},
\end{eqnarray}
where $ {}^{(3)}R$ is the Ricci curvature scalar for the three surface (in what follows we will drop the 3 superscript) and $\pi \equiv g_{ab} \pi^{ab}$.

At zeroth order, i.e. in the background Einstein static spacetime, there is only one nontrivial constraint namely
\begin{eqnarray}
\bar{\rho} &=& \frac{ 3 \Lambda }{2 \kappa }, 
\end{eqnarray}
where we have used the background conditions $\bar{R}_{ab} = \Lambda \bar{g}_{ab}$, $\bar{\pi}^{ab} = 0$.  At linear order (using the background equations) the 
constraint equations are 
\begin{eqnarray}
\delta R &=& 2 \kappa \delta \rho, \\
\nabladb \deltapiuab   &=& \kappa \sqrtgb \delta J^{a}, 
\end{eqnarray}
and similarly the second order equations are, using the linear and background equations,  
\begin{eqnarray}
\delta^{2} R &=& \frac{1}{| \gb |} \left( \deltapiuab \deltapidab - \frac{1}{2} \deltapitr^2 \right) + 2 \kappa \delta^{2} \rho \\
\nabladb \delta^{2}  \pi^{ab} 
+ 2 \left[ \delta C^{a}_{bm} \delta \pi^{mb}  + \delta C^{b}_{bm} \delta \pi^{am} \right]\sqrtgb  
										&=& \kappa \delta^{2} ( J^{a} \sqrt{ |g| } ), 
\end{eqnarray} 
Here $2 \delta C^{a}_{bm} \equiv \bar{g}^{a c} ( \nabladm \delta g_{bc} +  \nabladb \delta g_{mc} -  \nabladc \delta g_{bm} )$ is 
the perturbed connection. One can rewrite equation (19), the equation associated with time reparametrization invariance at second order, 
using  $\delta^{2} R = 2 \delta g^{ab} \delta R_{ab} + \bar{g}^{ab} \delta^{2} R_{ab} + \delta^{2} g^{ab} \bar{R}_{ab}$ to obtain
\begin{eqnarray}
\nonumber
\bar{ \Delta } \delta^{2} g - \nablaum \nablaub \delta^{2} g_{bm} - \Lambda \bar{ g }_{ab} \delta^{2} g^{ab} &=& \nabladell B^{\ell} 
					+ 2 \delta C^{c}_{b[a} \delta C^{\ell}_{\ell]c} \bar{g}^{ab} + 2 \delta g^{ab} \delta R_{ab}
- \frac{1 }{| \gb |} \left(  \delta \pi^{ab} \delta \pi_{ab} - \frac{1}{2} \delta \pi^{2} \right) - 2 \kappa \delta^{2} \rho, \\
\end{eqnarray}
where by $ \nabladell B^{\ell}$ we denote a combination of terms that occur in the form of a total derivative. In obtaining (21) we have expanded 
the second order Ricci scalar in terms of the perturbed connection defined above and grouped second order terms and products of linear order terms. 
One can similarly simplify the momentum constraints (20) but in this paper our main concern will be the second order timelike, or ``Hamiltonian'', constraint. 

Typically what one does is to solve the constraints for a given set of fluctuation variables and in that way solve for the constrained 
variables in terms of the free variables. A convenient way to facilitate this process is through an orthogonal decomposition of the 
fluctuations in $g_{ab}$ and $\pi^{ab}$ into longitudinal and transverse parts. As pointed out in \cite{Brill:1973} this procedure
differs crucially for tensors in closed spaces from the usual one in flat space since one can now have fluctuations in global parameters
such as the total volume of the space while still having no local excitations. 

Following the notation of Brill et al one may decompose the metric perturbation into its various transverse, transverse traceless, and longitudinal parts
via 
\begin{eqnarray}
\delta g_{ab} &=&   \delta g_{ab}^{(TT)} 
		+ \frac{1}{2} \left( - \nablada \nabladb  
+ \Deltabar \bar{g}_{ab} \right)\delta g^{(Tr)}  + \tilde{{\cal{A}}}_{\delta g}  \bar{g}_{ab} + 2 \delta g^{(V)}_{(a|b)}, 
\end{eqnarray}
where $\Deltabar \equiv \nabladc \nablauc$.
The transvserse traceless and longitudinal parts are defined as usual:
\begin{eqnarray}
 \bar{g}^{ab} \delta g_{ab}^{(TT)} &=& 0 = \nablaua  \delta g_{ab}^{(TT)}, \\
\nablada \delta {g^{(V)}}^{a} &=& 0, 
\end{eqnarray}
and here $\tilde{ {\cal{A}} }_{\delta g} \equiv {\cal{A}}_{\delta g} + \frac{\Lambda}{2}  \delta  {g^{(Tr)}}$.
The homogeneous fluctuations $\tilde{{\cal{A}}}_{\delta g}$ are essentially the global fluctuations (related 
to the volume fluctuation) unique to perturbations of closed spaces. An analagous decomposition holds for the momentum fluctuation (the
homogeneous modes $ \tilde{{\cal{A}}}_{\delta \pi} $ describe the `time rate of change` of the volume fluctuation). 

Using the above decomposition we note that an identity due to the symmetry of the background is that the perturbations are all `doubly transverse`. Indeed, we see
\begin{eqnarray}
\nablaub \delta g_{ab} &=& \bar{\Delta} \delta g_{a}^{(V)}   +   \bar{g}^{bm} \left( -\bar{R}^{\ell}_{\ bma} \delta g_{\ell}^{(V)} \right) 
									\stackrel{*}{=} ( \Deltabar + \Lambda ) \delta g_{a}^{(V)} \\
\nablaua \nablaub \delta g_{ab} &=& \bar{g}^{an} [ \nabladn, \Deltabar ] \delta g_{a}^{(V)} \stackrel{*}{=} 0  
\end{eqnarray}
We also note that we can write 
\begin{eqnarray}
\delta^{2} R &=& \Lambda \bar{g}_{ab} \delta^{2} g^{ab} + 2 \delta g^{ab} \delta R_{ab} 
+ 2 \bar{g}^{ab} \delta C^{c}_{b[a} \delta C^{\ell}_{\ell]c}  
				- 2 \bar{\nabla}_{[a} \delta^{2} \tilde{C}^{c}_{c]b} \bar{g}^{ab} + \nabladell B^{\ell},  
\end{eqnarray}
where
\begin{eqnarray}
2\delta^{2} \tilde{C}^{a}_{\ bc} &\equiv&  \bar{g}^{am} \left( \nabladb \delta^{2} g_{cm} 
					+  \nabladc \delta^{2} g_{bm} - \nabladm \delta^{2} g_{bc}  \right) 
\end{eqnarray}
We may use the decompositions above to eliminate the `doubly transverse` term and obtain
\begin{eqnarray}
\nonumber
 \Deltabar  \delta^{2} g 
-  \Lambda \bar{g}_{ab} \delta^{2} g^{ab} 
			&=&  2 \delta g^{ab} \delta R_{ab} + \nabladell B^{\ell} 
			     + 2 \delta C^{c}_{b[a} \delta C^{\ell}_{\ell]c} \bar{g}^{ab} \\
&&- \frac{1 }{| \gb |} \left(  \delta \pi^{ab} \delta \pi_{ab} - \frac{1}{2} \delta \pi^{2} \right) - 2 \kappa \delta^{2} \rho, 
\end{eqnarray}
where $\delta^{2} \rho$ are the second order energy density fluctuations. We explicitly insert the above decompositions into equation (28) in the next section.  

In the case of the generalized scalar field $\phi$ introduced in section II, the Hamiltonian formulation is more involved since the canonical momentum conjugate 
to $\phi$ is generally 
\begin{eqnarray}
\nonumber
\pi_{\phi} &\equiv& \frac{ \partial {\cal{L}} }{ \partial \dot{\phi}} = \frac{ \sqrt{| \gb | } }{{\cal{N}}} ( \dot{\phi} - {\cal{N}}^{i} \phi_{,i} ) 
		\left( \frac{ { \dot{\phi}}^{2} }{{\cal{N}}^2} - \frac{2 {\cal{N}}^{i} \dot{\phi} \phi_{,i} }{{\cal{N}}^2} - \phi^{,i} \phi_{,i} \right)^{\alpha - 1}, 
\end{eqnarray}
where $\cal{N}$ and ${\cal{N}}^{i}$ are the purely kinematical lapse and shift variables of the ADM formalism. For general $3 \ge \alpha \ge 1$ this expression cannot be 
inverted in 
closed form to find $\dot{\phi} = \dot{\phi}(\pi_{\phi})$, however for a spatially homogeneous scalar field it is possible and the resulting Hamiltonian (energy density) is 
\begin{eqnarray}
H_{0} &=& \frac{ 2 \alpha - 1}{2 \alpha } \left( \frac{ { ( \pi_{\phi} ) }^{2 \alpha} }{ \sqrt{ | g | } } \right)^{\frac{1}{2 \alpha -1}} + \frac{ \Lambda }{ \kappa }
      = \sqrt{ | g | } \rho, 
\end{eqnarray}
where $\bar{\pi}_{\phi} = \sqrt{ | \gb | } { \dot{\bar{\phi}}}^{2 \alpha -1 }$. We define the fluctuations 
$\delta \pi_{\phi}$ and $\delta^2 \pi_{\phi}$ using the formal definition of $\pi_{\phi}$ given above. For example the first order fluctuation in $\pi_{\phi}$ is formally 
defined by
\begin{eqnarray}
\delta \pi_{\phi} &\equiv& 
\frac{ \bar{\pi}_{\phi} }{ \sqrt{ |\gb| } } \left( \delta \left( \frac{ \sqrt{ |g| } }{ \cal{N} } \right) 
				+ \sqrtgb \left( \delta \dot{\phi } \frac{ 2 \alpha - 1}{ \dot{ \bar{ \phi } } } - 2( \alpha -1 ) \delta {\cal{N}} \right) \right), 
\end{eqnarray}
and similarly for the second order fluctuation (though new cross-terms like $\delta {\cal{N}}^{i} \delta \phi_{,i} \mbox{  and  } \delta \phi^{,i} \delta \phi_{,i}$ start 
to appear). 
Formally perturbing equations (11) and (12) with the above definitions of $\delta \pi_{\phi}$, $\delta^2 \pi_{\phi}$ yields 
\begin{eqnarray}
2 \kappa \delta \rho &=& \alpha \Lambda \frac{ \delta N }{ \bar{N} } 
				= \frac{2 \Lambda \alpha}{2 \alpha -1} \left[ \frac{ \delta \pi_{\phi} }{ \bar{\pi}_{\phi} } 
											- \frac{ \delta g}{2} \right], \\
2 \kappa \delta^{2} \rho &=& \alpha \Lambda \left[ (\alpha - 1) \left( \frac{\delta N}{ \bar{N}} \right)^2 + \frac{ \delta^{2} N }{ \bar{N} } \right], 
\end{eqnarray}
where the second order term is given explicitly by
\begin{eqnarray}
\nonumber
\delta^2 N  &=& 2 (  {\delta \dot{ \bar{ \phi } }}^2 + \dot{\bar{\phi}} \delta^2 \dot{\phi} )
			- 4 \phibdot \delta \dot{\phi} \delta \tilde{N} - 2 \phibdot^2 \delta^2 \tilde{N} + 6 \phibdot^2 ( \delta \tilde{N} )^2 
				- 2 \phibdot \delta \tilde{N}^{i} \delta \phi_{,i} - \delta \phi^{,i} \delta \phi_{,i}, 
\end{eqnarray}
so that, in Hamiltonian form, we finally obtain  
\begin{eqnarray}
2 \kappa \delta^2 \rho &=& \frac{2\alpha \Lambda}{2 \alpha -1} \left[ \frac{ \delta^{2} \pi_{\phi} }{ \bar{\pi}_{\phi} } - \frac{ \delta^{2} \sqrt{|g|} }{ \sqrt{ |\gb| } }
				- \frac{ \delta \phi^{,i} \delta \phi_{,i} }{2  } \left( \frac{ \bar{\pi}_{\phi} }{ \sqrt{ |\gb|} }  \right)^{\frac{2}{1-2\alpha} } \right . \\
\nonumber
&& \left . + \frac{1}{ (2\alpha -1)^2 } \left( (2\alpha^2 + 8\alpha -7) (\delta \ln \sqrt{|g|})^2 + (2\alpha^2+4\alpha-5) (\delta \ln \pi_{\phi} )^2
				   -4 (\alpha^2 + 3 \alpha -3) (\delta \ln \sqrt{|g|} )  (\delta \ln \pi_{\phi} ) \right) \right]
\end{eqnarray}
Here $\delta g \equiv \bar{g}^{ab} \delta g_{ab}$ and $\bar{N}^{\alpha} = \Lambda / (\kappa(2\alpha -1))$ by the zeroth order constraint (13).
Inserting the above matter perturbations into equation (25) and using the second-order identity 
$\bar{g}^{ab} \delta^{2} g_{ab} + \bar{g}_{ab} \delta^{2} g^{ab} = -2 \delta g^{ab} \delta g_{ab}$, we arrive at 
\begin{eqnarray}
\nonumber
( \Deltabar + \Lambda\frac{\alpha -1 }{2 \alpha -1}    )  \delta^{2} g
                            &=& \Lambda \left( \frac{2 - 3 \alpha}{2 \alpha -1} \right) \delta g^{ab} \delta g_{ab} + 2 \delta g^{ab} \delta R_{ab} 
			     + 2 \delta C^{c}_{b[a} \delta C^{\ell}_{\ell]c} \bar{g}^{ab}   
			- \frac{1 }{| \gb |} \left(  \delta \pi^{ab} \delta \pi_{ab} - \frac{1}{2} \delta \pi^{2} \right)  \\
\nonumber
&&  - \frac{2 \Lambda \alpha}{2 \alpha -1}\left[   \frac{ \delta^2 \pi_{\phi} }{ \bar{\pi}_{\phi} }  
				- \frac{ \delta \phi^{,i} \delta \phi_{,i} }{2  } \left( \frac{ \bar{\pi}_{\phi} }{ \sqrt{ |\gb|} } \right)^{\frac{2}{1-2\alpha} } 
 + \frac{1}{ (2\alpha -1)^2 }  \left( (-2\alpha^2 + 10\alpha -8) (\delta \ln \sqrt{|g|})^2 \right . \right . \\
&& \left. \left. + (2\alpha^2+4\alpha-5) (\delta \ln \pi_{\phi} )^2 -4 (\alpha^2 + 3 \alpha -3) (\delta \ln \sqrt{|g|} )  (\delta \ln \pi_{\phi} ) \right)
									 \right]  
\end{eqnarray}
The various terms in $( \delta \pi_{\phi} )$ can be reexpressed in terms of the metric fluctuations, using the linearized Hamiltonian constraint, via
\begin{eqnarray}
 \delta \ln \pi_{\phi} &=& 
	- \frac{ (2\alpha -1)  }{2 \Lambda \alpha } \left( \left[ \Deltabar +  \Lambda  \frac{ \alpha -1}{ 2 \alpha -1} \right] \delta g
			- \nablauc \nablaub \delta g_{cb}	\right) \equiv {\cal{M}} \delta g,
\end{eqnarray}
so that the only explicit matter dependence in the second order hamiltonian constraint appears through the $\delta ^{2} \pi_{\phi}$ and 
$\bar{g}^{ij} \delta \phi_{,i} \delta \phi_{,j}$ terms. 

\section{ Nonlinear restrictions on the linear fluctuations }

\subsection{ Fixing the linear gauge}
As is usual in relativistic perturbation theory one must fix the coordinate freedom inherent in the metric and matter fluctuations. In 
our particular case we want to remove the homogeneous second order matter dependence, which enters through the $\delta^{2} \pi_{\phi}$ term, in order 
that equation (35) is of the form ${\cal{L}} \delta^2 F = S ( (\delta F)^2 )$ (where ${\cal{L}}$ is an elliptic operator (with only constants in its kernel) 
acting on the fluctuations in some quantity $F$ in a closed space). In this form (since the timelike Killing vector component is trivial) the nonlinear constraints 
simply become $\int S = 0$ where the integral is over the closed space \cite{Brill:1973}. 

Fortunately the Einstein static background crucially simplifies the relevant gauge transformation laws of the fluctuations not only at linear order, 
but also at second order. Indeed the term $\delta^{2} \pi_{\phi}$ only depends on the {\it linear} gauge fixing essentially because 
$\dot{ \bar{\pi} }_{\phi} = 0$.  We are only interested in the homogeneous part of this $\delta^{2} \pi_{\phi}$ term so we pick a linear gauge-fixing to 
eliminate it and leave only the inhomogeneous part $\delta^2 \tilde{ \pi }_{\phi}$. 

We consider the linear spacetime coordinate transformation $\hat{x}^{\beta} = x^{\beta} + (T, \partial^{i}M + \tilde{M}^{i})$, where 
$(T,M, \tilde{M}^{i})$ satisfy
\begin{eqnarray}
\Deltabar \left( (2\alpha-1)  \Deltabar + (\alpha -1) \Lambda \right) T &=& -(2\alpha-1) \delta \pi^{(Tr)} \\
      \int_{S^3} \left\{ 2 M \Deltabar \delta \pi_{\phi} +   \delta^{2} \pi_{\phi} \right\} \sqrt{| \gb| } d^{3}x &=& 0  \\
\tilde{M}^{i} &=& - \delta g^{i}_{(V)},  \ \ \nabladi \tilde{M}^{i} = 0,  
\end{eqnarray}
and where the spatial dependence of the modes is understood in terms of the eigenfunctions of the spatial laplacian: 
$\Deltabar F = -\sum_{L} \frac{L(L+2)}{a_{0}^2} F = -  \sum_{L} \frac{ \Lambda \alpha}{\alpha + 1} L(L+2)F, L \in {\cal{Z}}^{+}$. 
In this new coordinate system $\delta g^{i}_{(V)} = 0$, the homogeneous part of $\delta^{2} \pi_{\phi}$ is zero, and $\delta \pi^{(Tr)}=0$. 
Thus
\begin{eqnarray}
- \frac{1 }{| \gb |} \left(  \delta \pi^{ab} \delta \pi_{ab} - \frac{1}{2} \delta \pi^{2} \right)
	-  \frac{2 \Lambda \alpha}{\bar{\pi}_{\phi} (2\alpha -1)} \delta^{2} \pi_{\phi} &=& - \frac{1 }{| \gb |}   \left( \delta \pi^{ab}_{TT} \delta \pi_{ab}^{TT} 
														- \frac{3}{2} (\tilde{\cal{A}}_{\delta \pi} )^2 \right)
										-  \frac{2 \Lambda \alpha}{\bar{\pi}_{\phi} (2\alpha -1)} \delta^{2} \tilde{\pi}_{\phi}.
\end{eqnarray}
This remarkable tranformation completely fixes the vector degrees of freedom at linear order, however we still have the freedom $M \rightarrow M + f(t)$ in the scalar sector
which we use to eliminate the homogeneous modes $\tilde{ {\cal{A}} }_{\delta g}$ by picking a special $f(t)$ (whose form is not particular illuminating at this stage).  
Furthermore, the linearized Hamiltonian constraint (36) in this coordinate system implies that $ \Deltabar \delta \pi_{\phi}$ cannot be zero everywhere, which means 
we can always pick a function $M$ such that equation (38) is satisfied.

\subsection{ Gauge-fixed nonlinear constraints }

The nonlinear constraint associated with (35) (an integrability condition on $\delta^{2} g^{(Tr)}$), is effectively the integral 
of the right hand side of equation (35) set to zero: 
\begin{eqnarray}
\nonumber
0 = \int_{S^{3}} &&  \Lambda \left( \frac{2 - 3 \alpha}{2 \alpha -1} \right) \delta g^{ab} \delta g_{ab} + 2 \delta g^{ab} \delta R_{ab} 
			     + 2 \delta C^{c}_{b[a} \delta C^{\ell}_{\ell]c} \bar{g}^{ab}   
			- \frac{1 }{| \gb |} \left(  \delta \pi^{ab} \delta \pi_{ab} - \frac{1}{2} \delta \pi^{2} \right)  \\
\nonumber
&&  - \frac{2 \Lambda \alpha}{2 \alpha -1}\left[   \frac{ \delta^2 \pi_{\phi} }{ \bar{\pi}_{\phi} }  
				- \frac{ \delta \phi^{,i} \delta \phi_{,i} }{2  } \left( \frac{ \bar{\pi}_{\phi} }{ \sqrt{ |\gb|} } \right)^{\frac{2}{1-2\alpha} } 
 + \frac{1}{ (2\alpha -1)^2 }  \left( (-2\alpha^2 + 10\alpha -8) (\delta \ln \sqrt{|g|})^2 \right . \right . \\
\nonumber
&& \left. \left. + (2\alpha^2+4\alpha-5) (\delta \ln \pi_{\phi} )^2 -4 (\alpha^2 + 3 \alpha -3) (\delta \ln \sqrt{|g|} )  (\delta \ln \pi_{\phi} ) \right)
									 \right]  \sqrtgb d^3x 
\end{eqnarray}
Using by-parts integration and compactness one can show that 
\begin{eqnarray}
\nonumber
2 \int_{S^{3}}  \sqrtgb \left(  \delta g^{ab} \delta R_{ab} +   \delta C^{c}_{b[a} \delta C^{\ell}_{\ell]c} \bar{g}^{ab} \right) &=& 
\int_{S^{3}}   \delta g^{ab} \delta R_{ab} \sqrtgb, 
\end{eqnarray}
which implies, using the linear equations of motion and making the above gauge choice,
\begin{eqnarray}
 \int_{S^{3}} &&  \frac{1}{ | \gb | }  \delta \pi^{ab}_{TT}  \delta \pi_{ab}^{TT} 
+  \delta g^{ab} \frac{\Deltabar  \delta g_{ab} }{2} - \frac{ \Lambda }{2(2 \alpha -1) } \delta g^{ab} \delta g_{ab}  \\
\nonumber
&&
 + \left( \frac{ 2 \Lambda \alpha }{ (2\alpha-1)^3 } \left( {\cal{M}}^2 (2 \alpha^2 + 4 \alpha - 5) - 2{\cal{M}} ( \alpha^2 + 3 \alpha -3) 
					+ \frac{ -2 \alpha^2 + 10 \alpha - 8}{4} \right) - \frac{ \Lambda }{2} \right) ( \delta g )^2
								 - 3\left( \frac{  { \tilde{ {\cal{A}} }_{\delta \pi} }^2 }{2} \right)   V = 0, 
\end{eqnarray}
where  $V$ represents the volume of the initial static space and ${\cal{M}}$ is defined in equation (31). We have eliminated the terms in $\delta \phi_{,i}$ by 
using the scalar-sector momentum constraints (which are $\partial^{i}(\frac{ \Lambda }{2 \kappa \bar{\pi}_{\phi} } \delta \pi^{(Tr)} - \delta \phi )=0$ ).  The constraint 
is split into its inhomogeneous ($L \ge 2$) and homogeneous ($L = 0$) pieces. \footnote{The fluctuations corresponding to $L = 1$ can be shown to be purely coordinate 
fluctuations.}. 

Finally, inserting the decompositions into equation (41) and using equation (36) to remove the dependence in $( \delta \pi_{\phi} )^2$, we arrive at  
\begin{eqnarray}
0 &=& \int_{S^{3}}  A \sqrt{ |\gb| } d^{3}x  -  3\left( \frac{  { \tilde{ {\cal{A}} }_{\delta \pi} }^2 }{2} \right) V  , 
\end{eqnarray}
where
\begin{eqnarray}
 A &=&  \sum_{L \ge 2} \left[ \alpha \Lambda^3  k^2  
	\left\{   \frac{\alpha^2 (2\alpha^2+4\alpha-5)}{2(\alpha+1)^4 (2\alpha-1)} k^6
		- \frac{\alpha (6 \alpha^2 + 21 \alpha -20)}{4 (2\alpha-1)(\alpha+1)^3 }k^4 
		+ \frac{100 \alpha^3 - 194 \alpha^2 + 109 \alpha + 8 \alpha^2 - 20}{8(\alpha+1)^3 (2\alpha-1)^3}k^2 \right . \right . \\
\nonumber
&&
\left . \left . + \frac{\alpha - 1}{2(2\alpha-1)(\alpha+1)} \right\} ( \delta g^{(tr)} )^2  + \left\{ \frac{1}{| \gb | } 
\left[  \delta \pi_{ab}^{(TT)}  \delta \pi_{\ell m}^{(TT)} \right] 
+ \frac{ \Lambda }{2\alpha + 1} ( \alpha(k^2 - 2) + \frac{ \alpha + 1}{2\alpha -1}   ) 
\delta g_{\ell m}^{(TT)} \delta g_{ab}^{(TT)} \right\} \bar{g}^{\ell a} \bar{g}^{mb} \right]
\end{eqnarray}
and where $k^2 \equiv L(L+2)$. We point out that equation (42) is not the integral of the second order Hamiltonian $H^{(2)}$ 
for the fluctuations since, given our gauge fixing, the symplectic terms $\delta \pi^{(Tr)} \delta \dot{g}_{Tr}^{ij}$ do not contribute.

The main result of this paper is that first term in equation (43), $A$, is positive definite given $\alpha \ge 1$, $L \ge2$. We observe that the global constraint is an integral 
over $S^3$, i.e. is of positive measure. This means that in the absence 
of the homogeneous mode, which provide a strictly negative definite contribution to the integral through the second term, there is no nontrivial solution 
to these global constraints {\it even though} there are certainly solutions to the second order equations which only have inhomogeneous linear seeds. 
In other words, if one wants to study the evolution of the second order modes one must include, as part of their source, the zero mode at linear order in 
order to properly satisfy the initial value constraints. Therefore, we have shown that at linear order we {\it must} include the  unstable, homogeneous mode. We 
emphasize that we are {\it not} claiming stability or not at the second order level, rather we are claiming instability at the {\it linear} level. Whether or 
not second order perturbations could stabilize the spacetime at a sufficiently high value of the linearized solution remains unclear, but we do not address that issue 
here though we strongly suspect not. 

\section { Conclusions }

We have shown that the requirement that the second order Einstein
constraint equations be integrable demands that any non-homogenous
linear mode perturbations of the einstein static universe  must be accompanied
 by the homogenous linear mode with comparable amplitude. Since this
homogeneous linear mode is exponentially unstable, this implies that
any linear approximation to a solution of Einstein's equations must be
unstable. Our result is valid for perfect fluid matter determined by a
potential and with a constant velocity of sound.


\section{ Acknolwedgements }

Both authors thank NSERC for support during the completion of this work. W.G.U. also thanks the CIAR for support.

\end{document}